      [1999/12/01 v1.4c Il Nuovo Cimento]
\documentclass{cimento}
\usepackage{graphicx}  

\newcommand{\begit}{\begin{itemize}}
\newcommand{\enit}{\end{itemize}}
\newcommand{\begen}{\begin{enumerate}}
\newcommand{\enen}{\end{enumerate}}
\newcommand{\beq}{\begin{equation}} 
\newcommand{\eeq}{\end{equation}} 
\newcommand{\beqa}{\begin{eqnarray}} 
\newcommand{\eeqa}{\end{eqnarray}}

\newcommand{\bof}{B_{0_{15}}}

\newcommand{\rnuten}{R_{\nu_{10}}}

\newcommand{\mdotth}{\dot{M}_{-3}}
\newcommand{\mns}{M_{1.4}}
\def\lesssim{\mathrel{\hbox{\rlap{\hbox{\lower4pt\hbox{$\sim$}}}\hbox{$<$}}}}
\def\gtrsim{\mathrel{\hbox{\rlap{\hbox{\lower4pt\hbox{$\sim$}}}\hbox{$>$}}}}
\title{Millisecond Proto-Magnetars \& Gamma Ray Bursts}
\author{Todd A.~Thompson\from{ins:x}\thanks{Hubble fellow.}}
\instlist{\inst{ins:x} Department of Astronomy \& Theoretical
Astrophysics Center, 601 Campbell Hall, The University of California, Berkeley, 94720}
\begin{document}

\maketitle

\begin{abstract}
In the seconds after core collapse and explosion, a thermal neutrino-driven
wind emerges from the cooling, deleptonizing newly-born neutron star.  
If the neutron star has a large-scale magnetar-strength surface magnetic
field and millisecond rotation period, then the wind is
driven primarily by magneto-centrifugal slinging, and only secondarily by
neutrino interactions.  The strong magnetic field forces the wind to corotate
with the stellar surface and the neutron star's rotational energy is efficiently extracted.
As the neutron star cools, and the wind becomes increasingly magnetically-dominated,
the outflow becomes relativistic.  
Here I review the millisecond magnetar model for long-duration gamma ray 
bursts and explore some of the basic physics of neutrino-magnetocentrifugal winds.
I further speculate on some issues of collimation and geometry in the millisecond
magnetar model.
\end{abstract}

\vspace*{-.75cm}
\section{Introduction}

A successful core-collapse supernova (SN) leaves behind a hot deleptonizing
proto-neutron star (PNS) that cools and contracts on its Kelvin-Helmholtz cooling 
timescale ($\tau_{\rm KH}\sim10$ s), 
radiating its gravitational binding energy ($\sim10^{53}$ ergs) in neutrinos
\cite{burrows_lattimer}.  
A small  fraction of these neutrinos deposit their energy in the tenuous atmosphere of the
PNS through the interactions $\nu_e n\rightarrow p e^-$, 
$\bar{\nu}_e p\rightarrow n e^+$, $\nu\bar{\nu}\rightarrow e^+e^-$,
and $\nu (e^-e^+)\rightarrow\nu^\prime (e^{-\prime}e^{+\prime})$.  
Inverse processes provide cooling.  Net neutrino heating drives a thermal 
wind that emerges into the post-supernova shock environment \cite{bhf}.
For typical non-rotating non-magnetic (NRNM)
neutron stars, the total kinetic energy 
of the wind over $\tau_{\rm KH}$ is of order $\sim10^{48}$ ergs and, hence, the 
addition to the asymptotic SN energetics is small on the scale of the canonical 
SN energy, $E_{\rm SN}\sim10^{51}$ ergs \cite{dsw,qw,tbm}.  

Magnetars --- a class of neutron stars with 
surface magnetic field strengths of $B_0\sim10^{15}$ G ---
are thought to be born with millisecond rotation periods,
their intense fields having been generated by an efficient dynamo 
\cite{duncan_thompson,thompson_duncan,kouv}.
Millisecond rotation periods imply a reservoir of rotational energy that is large on the 
scale of $E_{\rm SN}$: $E_{\rm Rot}\sim2\times10^{52}\,\,{\rm ergs}\,\,\,
\mns\,\rnuten^2\,P_1^{-2}$,
where $M_{1.4}=M/1.4$ M$_\odot$, $R_{\nu_{10}}=R_\nu/10\,\,{\rm \,km}$, 
and $P_1$ is the spin period in units of 1 millisecond (ms).   Stellar progenitors
that produce millisecond magnetars (MSMs) must have iron cores rotating with periods of
$\lesssim10$ s at the moment of collapse \cite{hlw}.   
The character of collapse, bounce, and explosion
can be modified by rotation.  For example, rotational
support leads to characteristically lower neutrino luminosity ($L_\nu$) and 
longer $\tau_{\rm KH}$ \cite{fryer_heger,tqb}.  In addition, a fraction of 
the gravitational binding of rotating collapse may be stored in shear energy and
tapped by viscous processes; for parameters appropriate to MSM birth this effect
can yield explosions in models  that would otherwise fail \cite{tqb}.  Although rotation
may be important during collapse and explosion,
and although small-scale
magnetic fields may be important in providing a viscosity capable of tapping
the shear energy generated during collapse, even the magnetic energy density associated with
a dipole field strength of $10^{15}$ G is small with respect to the thermal pressure
exterior to any PNS in the first $\sim1-2$ seconds after collapse.
Therefore, similar to NRNM PNSs, we expect that the wind that accompanies MSM cooling
is driven by neutrino heating at early times.
However, as $L_\nu$ decreases and the thermal pressure exterior
to the MSM decreases, the region exterior to the MSM must 
become magnetically-dominated. 

The strong magnetic field forces the matter composing 
the outflow into near corotation with the stellar surface out to $\sim R_{\rm A}$, 
the Alfv\'{e}n point, where the magnetic energy density equals the kinetic energy 
density of the outflow.  For $P\sim1$ ms, if $R_{\rm A}\gtrsim15$ km, then the
wind is driven primarily by magneto-centrifugal slinging; neutrino heating becomes
relatively unimportant in determining the asymptotic wind velocity.
Rotational energy is transferred
directly from the MSM to the wind and this provides an efficient mechanism for spindown
\cite{schatzman,mestel_spruit}.  The spin period of the MSM $e$-folds on 
the spindown timescale $\tau_{\rm J}\approx(2/5)(M/\dot{M})(R_\nu/R_{\rm A})^2$,
where $\dot{M}$ is the mass loss rate.
Because $E_{\rm Rot}\gg E_{\rm SN}$, just one $e$-folding of $\Omega$ is sufficient
to modify the dynamics of the SN remnant significantly.  
If $\tau_{\rm J}$ is small with respect to the
time for the SN shock to traverse the progenitor ($\sim$ tens of seconds for type-Ib, -Ic
progenitors) we also expect this extra energy source to modify the SN nucleosynthesis \cite{tcq}.
As the MSM cools and the outflow becomes increasingly magnetically-dominated, $R_{\rm A}$
increases.  It cannot do so indefinitely.  $R_{\rm A}$ approaches the radius of the light cylinder
$R_{\rm L}=c/\Omega\simeq48 P_1$ km asymptotically.
As it does so, the flow becomes increasingly relativistic.
This is the transition between non-relativistic mass-loaded outflow and relativistic
Poynting-flux dominated neutron star wind.  
All neutron stars, regardless of their initial spin period and magnetic field strength,
go through this transition.  MSMs are interesting because
this transition occurs at high wind kinetic luminosity.  Because $E_{\rm Rot}$
is large and the spindown timescale is short, and because the velocity of the wind must
eventually become relativistic, these objects are a natural candidate for the central engines
of long-duration gamma ray bursts (GRBs) \cite{usov,thompson94,tcq}.  

\section{Proto-Magnetar Spindown}

In this section I summarize the results of Ref.~\cite{tcq}. See that work for more details.

Angular momentum conservation implies that 
$\dot{J}=d/dt(I\Omega)=-\dot{M}{\cal L}$, where $\cal{L}$ is the specific angular momentum
carried by the wind and $I$ is the moment of inertia. 
In the classic model for solar spindown constructed by 
Ref.~\cite{weber_davis}, the wind problem is treated in one spatial dimension and
in the equatorial plane. Consideration of the azimuthal momentum 
equation together with Faraday's law gives ${\cal L}=R_{\rm A}^2\Omega$.
To estimate the angular momentum loss rate and, thereby, the wind luminosity and
the spindown timescale, we must first estimate $R_{\rm A}$.  The location of the
Alfv\'en point depends on the radial dependence of the poloidal magnetic field.
Because models with purely monopole fields generally over-estimate spindown and
models with pure dipole fields under-estimate spindown, we parameterize 
$B_r=B_0(R_\nu/r)^{\eta}$,  where $2\lesssim\eta\lesssim3$.
Taking $B^2/8\pi\sim\rho v_r^2/2$ at $R_{\rm A}$, 
$\rho=\dot{M}/4\pi r^2 v_r$, and assuming that 
$v_{\rm A}=v_r(R_{\rm A})\sim v_\phi(R_{\rm A})\sim R_{\rm A}\Omega$, we find that 
\beq
R_A=B_0^{2/(2\eta-1)}\,R_\nu^{2\eta/(2\eta-1)}\,(\dot{M}\Omega)^{-1/(2\eta-1)},
\label{slowspin2}
\eeq
where $v_{\rm A}$ is the radial Alfv\'{e}n speed, $v_r$ is the radial velocity, $v_\phi$
is the azimuthal velocity, $\rho$ is the mass density, and $B_r$ is the radial magnetic field.
Equation (\ref{slowspin2}) assumes that $R_{\rm A}\Omega\gg v_\nu$, where 
$v_\nu$ is the
asymptotic wind velocity in a NRNM outflow 
($v_\nu\lesssim3\times10^9$ cm s$^{-1}$).
The absolute value of the rotational energy loss rate 
can be written as  
\beq
\dot{E}_{\rm NR}=
B_0^{4/(2\eta-1)}R_\nu^{4\eta/(2\eta-1)}
\dot{M}^{(2\eta-3)/(2\eta-1)}\Omega^{(4\eta-4)/(2\eta-1)}.
\eeq
For parameters appropriate to MSMs
\beqa
\dot{E}_{\rm NR}^{\eta=2}
&\simeq&
1.5\times10^{51}\bof^{4/3}\rnuten^{8/3}\mdotth^{1/3} P_1^{-4/3}
\,\,\,{\rm ergs\,\,s^{-1}}
 \label{nonreletot1} \\
\dot{E}_{\rm NR}^{\eta=3}
&\simeq&4.5\times10^{50}\bof^{4/5}\rnuten^{12/5}\mdotth^{3/5} P_1^{-8/5} 
\,\,\,{\rm ergs\,\,s^{-1}}.
\label{nonreletot}
\eeqa
The subscript `NR' is added to emphasize that when the flow is non-relativistic,
$\dot{E}$ depends explicitly on $\dot{M}$.

The non-relativistic scalings for the energy loss rate can be compared with those in
the relativistic regime.  As $R_{\rm A}$ becomes close to $R_{\rm L}$ as $\dot{M}$
decreases during the cooling epoch, $v_{\rm A}$ approaches $c$ and the flow
becomes relativistic.
At $R_{\rm L}$, the ratio of magnetic to kinetic energy density is 
\beq
\Gamma=\left. \frac{B^2}{4\pi\rho c^2}\right|_{R_{\rm L}}=\,\,\,
B_0^2 R_\nu^{2\eta}\Omega^{2\eta-2}c^{1-2\eta}\dot{M}^{-1}
\label{gammalim}
 \eeq
and the energy loss rate is
\beq
\dot{E}_{\rm R}=-\Gamma\dot{M}c^2= -B_0^2 R_\nu^{2\eta}
\Omega^{2\eta-2}c^{3-2\eta}.
\label{reletot}
\eeq 
For $\eta=3$ the classical ``vacuum dipole'' limit is obtained; 
$\dot{E}_{\rm R}(\eta=3)=
B_0^2R_\nu^6\Omega^4c^{-3}$.\footnote{In the true ``vacuum dipole'' limit 
$\dot{E}_{\rm R}$ has a term $\sin\alpha/6$, where $\alpha$ is the 
angle between the spin axis and the magnetic dipole axis.}
For the monopole case with $\eta=2$, $\dot{E}_{\rm R}$ is larger 
than the dipole limit by a factor of $c^2/(\Omega R_\nu)^2$ --- 
a factor of $\sim23$ for a 10 km MSM with a 1 ms 
spin period.

The non-relativistic spindown rate
is larger than the relativistic spindown rate as a result of 
mass-loading.
To see this explicitly, note that 
\beqa
\dot{E}_{\rm NR}^{\eta=2}/\dot{E}_{\rm R}^{\eta=2}
&\simeq&1\,\bof^{-2/3}\rnuten^{-4/3}\mdotth^{1/3}P_1^{2/3} \\
\dot{E}_{\rm NR}^{\eta=3}/\dot{E}_{\rm R}^{\eta=3}
&\simeq&8\,\bof^{-6/5}\rnuten^{-18/5}\mdotth^{3/5}P_1^{12/5}.
\eeqa
For $\eta=2$ we see that the ratio is approximately unity, reflecting the fact that
for the parameters chosen $R_{\rm A}\sim R_{\rm L}$.  For slower spin periods,
the ratio increases, but not dramatically.  In contrast, the ratio of 
$\dot{E}_{\rm NR}$ to $\dot{E}_{\rm R}$ for the dipole case ($\eta=3$) 
is large for MSMs and it has a strong dependence on $P$.
Thus, for a magnetar born with a 10 ms spin period a naive application of the ``vacuum 
dipole'' formula underestimates the magnitude of the rotational energy loss rate by a factor
of $\sim2000$.  More detailed calculations reveal that when $P\sim10$ ms, for modest
neutrino luminosities, $\dot{M}$ is probably closer to $\sim10^{-5}$ M$_\odot$ s$^{-1}$
so that the ratio of $\dot{E}_{\rm NR}$ to $\dot{E}_{\rm R}$ is closer to $\sim120$ than
2000.  Even so, $\dot{E}_{\rm NR}/\dot{E}_{\rm R}$ is very large and 
an application of the relativistic formula in an epoch when $\dot{M}$ is large is incorrect.
Of course, if the field structure is not purely dipolar (that is, $\eta<3$), then
the discrepancy between the ``vacuum dipole'' spindown approximation and the
``true'' spindown rate becomes even larger.  In this case we would compare
$\dot{E}_{\rm NR}^{\eta=2}/\dot{E}_{\rm R}^{\eta=3}\simeq25\bof^{-2/3}
\rnuten^{-10/3}\mdotth^{1/3}P_1^{8/3}$, a factor of 25 for a MSM.
For a magnetar with a 10 ms spin period and lower mass loss rate the ratio becomes
$\dot{E}_{\rm NR}^{\eta=2}/\dot{E}_{\rm R}^{\eta=3}
\simeq2500\bof^{-2/3}\rnuten^{-10/3}\dot{M}_{-5}^{1/3}P_{10}^{8/3}$.
These arguments serve to underscore the fact that in the very early stages of proto-magnetar cooling,
the multi-dimensional structure of the wind/magnetic field interaction must be solved consistently
(to determine the effective $\eta$) 
and that inferences about the ``initial'' spin period of magnetars should not be based
on an application of the vacuum dipole approximation when $\dot{M}$ is large. 

For fixed $B_0$, $R_\nu$, and $\Omega$ the transition from non-relativistic (eq.~[\ref{nonreletot}])
to relativistic (eq.~[\ref{reletot}]) outflow occurs when $R_{\rm A}\sim R_{\rm L}$ 
and this point in time corresponds to a critical mass loss rate
$\dot{M}_{\rm crit}=B_0^2R_\nu^{2\eta}\Omega^{2\eta-2}c^{1-2\eta}$,
which scales with $P^{-2}$ and $P^{-4}$ for $\eta=2$ and $\eta=3$,
respectively. Because $\dot{M}$ scales with the neutrino luminosity and because --- to first
approximation --- the luminosity is a monotonically decreasing function of time, this scaling of 
$\dot{M}_{\rm crit}$ with $P$ implies that the wind is non-relativistic
for a larger fraction of the cooling time for longer rotation periods: for $\eta=3$ and $P=10$ ms
$\dot{M}_{\rm crit}\simeq3\times10^{-9}$ M$_\odot$ s$^{-1}$. Such a low $\dot{M}$
may correspond to a time several tens of seconds after collapse.


\section{Are Millisecond Proto-Magnetars GRB Central Engines?}

The spindown timescale $\Omega/\dot{\Omega}$ in the non-relativistic limit is 
\beqa
\tau_{\rm J_{NR}}^{\eta=2} &\simeq& 30\,\,{\rm s}\,\,\,
\mns\,\mdotth^{-1/3}\,\rnuten^{-2/3}\,\bof^{-4/3}\,P_{1}^{-2/3}
\label{slowspinscale2} \\
\tau_{\rm J_{NR}}^{\eta=3} &\simeq&96\,\,{\rm s}\,\,\,
\mns\,\mdotth^{-3/5}\,\rnuten^{-2/5}\,\bof^{-4/5}\,P_{1}^{-2/5}
\label{slowspinscale3}
\eeqa
For a MSM with $2\times10^{52}$ ergs of rotational energy we need only wait a 
fraction of $\tau_{\rm J}$ to extract an amount of energy comparable to
the supernova energy, $\sim10^{51}$ ergs.
Because $\tau_{\rm J}$ (or $\sim\tau_{\rm J}/5$) is 
comparable to $\tau_{\rm KH}$ we infer that significant
spindown may occur during the cooling epoch.  
As implied by the discussion of $\dot{E}_{\rm NR}$ above, 
for larger initial spin periods the spindown timescales {\it decrease}. 
In the relativistic limit ($R_{\rm A}\sim R_{\rm L}$)
\beqa
\tau_{\rm J_{R}}^{\eta=2} &\simeq& 34\,\,{\rm s}\,\,\,
\mns\,\rnuten^{-2}\,\bof^{-2}
\label{relspinscale2} \\
\tau_{\rm J_{R}}^{\eta=3} &\simeq&760\,\,{\rm s}\,\,\,
\mns\,\rnuten^{-4}\,\bof^{-2}\,P_{1}^{2}
\label{relspinscale3}
\eeqa
Although $\tau_{\rm J_{R}}^{\eta=2}$ does not depend on $\Omega$ explicitly,
it  may have an implicit dependence on
$\Omega$ if the magnetic field is generated by a 
dynamo \cite{duncan_thompson,thompson_duncan}.

With these timescales in hand we can (in a rudimentary way)
attempt to assess the MSM GRB mechanism.  A more detailed assessment
must await a consistent multi-d MHD solution.  From eqs.~(\ref{nonreletot1}),
(\ref{nonreletot}), (\ref{slowspinscale2}), and
(\ref{slowspinscale3}) we see that in the non-relativistic limit, 
regardless of $\eta$, the amount of energy 
extractable on $\sim10-100$ second timescales is in the range appropriate to
GRBs.  For $\eta<3$ these conclusions are stronger. Because $\gtrsim10^{51}$ ergs
can be extracted on a timescale shorter than or comparable to the timescale for the SN
shock to traverse the progenitor, we expect that the wind may 
significantly affect the dynamics of the remnant and the  $^{56}$Ni yield. In this way
it may be possible to generate hyper-energetic or 1998bw-like SNe during
MSM birth \cite{tcq,woosley_heger}.  The inferred energetics and $^{56}$Ni yield of SN2003dh
and SN1998bw put strong constraints on any GRB mechanism.  In the collapsar model
a disk wind is thought to generate the $^{56}$Ni required to power the SN lightcurve
\cite{macfadyen_woosley,pth}.  In the millisecond magnetar model, the energetic
wind shocks the material already processed by the supernova shock, perhaps generating
the large inferred $^{56}$Ni yields;  we are currently investigating the timing 
of this scenario.

Although it seems possible that MSM winds may generate energetic winds at early times,
the flow during this mass-loaded wind phase --- at least, on average --- is not relativistic. 
It is possible that a strong latitudinal dependence to the mass loss rate may yield
relativistic asymptotic velocities for matter emerging from mid-latitudes even when
our estimates would indicate $R_{\rm A}<R_{\rm L}$, but such a speculation must
be tested against realistic multi-d models.  It is also possible that large
temporal variations in $\dot{M}$ could cause the wind to alternate rapidly between 
non-relativistic and relativistic.   Strong variations in the mass loading 
could be caused by shearing of large-scale closed magnetic loops on the surface of the
fully convective MSM core \cite{thompson94}.   

As the flow becomes increasingly relativistic (on average),
we see from equation (\ref{relspinscale3}) that if the relativistic dipole limit strictly 
obtains, then the spindown timescale is long for $B\sim10^{15}$ G 
and $P\sim1$ ms.  Although relativistic spindown with $\eta=3$ will affect the asymptotic 
remnant dynamics by injecting energy over a long timescale,  it will probably
not generate a GRB with duration $\sim30$ seconds and energy $\sim10^{51}$ ergs.
Based on the scalings derived here, higher magnetic field strength, shorter spin period, 
or $\eta<3$ is probably required in order for MSM spindown to power GRBs.
If one of these possibilities obtains, then from the estimates 
in Ref.~\cite{tcq} we find that essentially all of the magnetic energy
at $R_{\rm L}$ must be transferred to the wind in order for the flow to obtain high
asymptotic Lorentz factor with large enough $\dot{E}$.  This is presumably accomplished 
by magnetic dissipation \cite{drenk,bland}. 

\section{Emergence, Geometry, \& Collimation}

If a relativistic outflow with the requisite energy to power a GRB can be generated by a MSM, it must emerge 
from the massive stellar progenitor.  The highly energetic non-relativistic wind, which precedes the relativistic
outflow, will likely be collimated by hoop stress and will therefore shape the cavity into which
the relativistic wind emerges.  Because the non-relativistic wind carries little mass in comparison 
with the overlying star, the relativistic outflow is not additionally hindered in its escape
from the progenitor by the preceding slow wind.  Hence, 
if the relativistic outflow can be collimated, then the dynamics of its emergence from the progenitor 
should be qualitatively similar to models of collapsar jets escaping Type-Ibc progenitors \cite{zhang}. 

One important possible objection to the MSM mechanism for GRBs is that it is difficult to 
collimate relativistic Poynting-flux dominated outflows \cite{begelman_li}.  
Observational evidence
for collimation in GRBs is abundant and so, at face-value, this would seem to be a problem.
There are at least three responses to this objection.  The first possible response is that the interaction
between the emerging and energetic (non-relativistic and then relativistic) 
wind with the overlying post-supernova-explosion ejecta may act to collimate the outflow.  
Future multi-d simulations should address this issue in detail.
The second potential response is that, in fact, relativistic  Poynting-flux dominated winds {\it can}
be efficiently collimated, as in the work of Ref.~\cite{konigl1}.
A final possible response is this: 
given the basic geometry of the magnetocentrifugal wind, it seems natural to suppose
that the asymptotic radial velocity is largest in the equatorial region.  This follows
both from the fact that centrifugal acceleration is largest at the equator and 
that the equatorial current sheet may facilitate significant dissipation of the 
magnetic energy.  Pulsars (e.g., the Crab) provide 
evidence for high Lorentz factor and energetically-dominant equatorial winds 
\cite{komissarov,spitkovsky}.
In analogy, is it possible that the geometry of GRBs is ``sheet"-like
(equatorial) rather than jet-like?  The solid angle subtended by a sheet with opening 
angle $\theta$ is $\sim\theta$,
whereas for a jet it is $\sim\theta^2$.  
This fact has implications for the afterglow --- 
a sheet-break rather than a jet-break \cite{rhoads,sari} --- and for the GRB census.  
Inferences from detailed predictions of the flux evolution may be used to rule out or
confirm the possibility of a sheet-like geometry in some bursts.  
I am constructing such models now.

\acknowledgments
I thank Philip Chang and Eliot Quataert for collaboration
and Jon Arons, Brian Metzger, Niccolo Bucciantini,
and Anatoly Spitkovsky for helpful conversations.
Support for this work is provided by NASA through Hubble Fellowship
grant \#HST-HF-01157.01-A awarded by the Space Telescope Science
Institute, which is operated by the Association of Universities for 
Research in Astronomy, Inc., for NASA, under contract NAS 5-26555.


\vspace*{-1cm}

\begin{thebibliography}{0}

\bibitem{burrows_lattimer}
\BY{Burrows, A. \& Lattimer, J.~M.}
\IN{Astrophys. J.}{307}{1986}{178} 

\bibitem{bhf}
\BY{Burrows, A., Hayes, J., \& Fryxell, B. A.}
\IN{Astrophys. J.}{450}{1995}{830}

\bibitem{dsw}
\BY{Duncan, R.~C., Shapiro, S.~L., \& Wasserman, I.}
\IN{Astrophys. J.}{309}{1986}{141} 

\bibitem{qw}
\BY{Qian, Y.-Z. \& Woosley, S.~E.} 
\IN{Astrophys. J.}{471}{1996}{331}

\bibitem{tbm}
\BY{Thompson, T.~A., Burrows, A., \& Meyer, B.}
\IN{Astrophys. J.}{562}{2001}{887}

\bibitem{duncan_thompson}
\BY{Duncan, R.~C.~\& Thompson, C.}
\IN{Astrophys. J. Lett.}{392}{1992}{9}

\bibitem{thompson_duncan}
\BY{Thompson, C.~\& Duncan, R.~C.}
\IN{Astrophys. J.}{408}{1993}{194}

\bibitem{kouv}
\BY{Kouveliotou, C.~et al.}
\IN{Astrophys. J. Lett.}{510}{1999}{115}

\bibitem{hlw}
\BY{Heger, A., Langer, N., Woosley, S.~E.}
\IN{Astrophys. J.}{528}{2000}{368}

\bibitem{fryer_heger}
\BY{Fryer, C.~L. \& Heger, A.}
\IN{Astrophys. J.}{541}{2000}{1033}

\bibitem{tqb}
\BY{Thompson, T.~A., Quataert, E., \& Burrows, A.}
\IN{Astrophys. J.}{620}{2005}{861}

\bibitem{tcq}
\BY{Thompson, T.~A., Chang, P., \& Quataert, E.}
\IN{Astrophys. J.}{611}{2004}{380}

\bibitem{schatzman}
\BY{Schatzman, E.}
\IN{Annales d'Astrophysique}{25}{1962}{18}

\bibitem{mestel_spruit}
\BY{Mestel, L.~\& Spruit, H.~C.}
\IN{Mon.~Not.~R.~Astron.~Soc.}{226}{1987}{57}

\bibitem{usov}
\BY{Usov, V.~V.}
\IN{Nature}{357}{1992}{472}

\bibitem{thompson94}
\BY{Thompson, C.}
\IN{Mon.~Not.~R.~Astron.~Soc.}{270}{1994}{480}

\bibitem{weber_davis}
\BY{Weber, E.~J.~\& Davis, L.}
\IN{Astrophys. J.}{148}{1967}{217}

\bibitem{woosley_heger}
\BY{Woosley, S.~E.~\& Heger, A.}
\IN{submitted to Astrophys. J.}{}{2003}{astro-ph/0309165}

\bibitem{macfadyen_woosley} 
\BY{MacFadyen, A.~I. \& Woosley, S.~E.}
\IN{Astrophys. J.}{524}{1999}{262}

\bibitem{pth}
\BY{Pruet, J., Thompson, T.~A., Hoffman, R.}
\IN{Astrophys. J.}{606}{2004}{1006}

\bibitem{drenk}
\BY{Drenkhahn, G.~\& Spruit, H.~C.}
\IN{Astro. \& Astrophys.}{391}{2002}{1141}

\bibitem{bland}
\BY{Lyutikov, M.~\& Blandford, R.~D.}
\IN{astro-ph/0312347}{}{2003}{}

\bibitem{zhang}
\BY{Zhang, W., Woosley, S.~E., \& Heger, A.}
\IN{Astophys. J.}{608}{2004}{365}

\bibitem{begelman_li}
\BY{Begelman, M.~C. \& Li, Z.-Y.}
\IN{Astrophys. J.}{426}{1994}{269}

\bibitem{konigl1}
\BY{Vlahakis, N.~\& Konigl, A.}
\IN{Astrophys. J.}{596}{2003}{1080}

\bibitem{komissarov} 
\BY{Komissarov, S.~S., \& Lyubarsky, Y.~E.}
\IN{Mon. Not. R. Astron. Soc.}{349}{2004}{779}

\bibitem{spitkovsky} 
\BY{Spitkovsky, A., \& Arons, J.}
\IN{Astrophys. J.}{603}{2004}{669}

\bibitem{rhoads}
\BY{Rhoads, J.~E.}
\IN{Astrophys. J.}{525}{1999}{737}

\bibitem{sari}
\BY{Sari, R., Piran, T., \& Halpern, J.~P.}
\IN{Astrophys. J. Lett.}{519}{1999}{L17}

\end{thebibliography}
\end{document}